\documentclass[aps,twocolumn,prb,showpacs,balancelastpage,amssymb,groupedaddress]{revtex4}
\usepackage{epsfig}
\usepackage{bm}

\begin{document}

\title{ Effects of relative orientation of the molecules on electron transport in molecular devices }

\author{ZHOU Yan-Hong$^{1}$}

\author{ ZHENG Xiao-Hong$^{2}$}

\author{ XU Ying$^1$}

\author{ ZENG Zhao-Yang$^{1}$}
\email{zyzeng@jxnu.edu.cn}

\author{ ZENG Zhi$^{2}$}

\affiliation { $^1$Department of Physics, Jangxi Normal University, Nanchang 330022, China\\
$^2$Key Laboratory of Material Physics, Institute of Solid State
Physics, Chinese Academy of Sciences, Hefei, Anhui 230031, China}

\date{May 29 2006}

\begin{abstract}

Effects of relative orientation of the molecules on electron
transport in molecular devices are studied by non-equilibrium
Green's function method based on density functional theory. In
particular, two molecular devices, with the planer Au$_{7}$ and
Ag$_{3}$ clusters sandwiched between the Al(100) electrodes are
studied. In each device, two typical configurations with the
clusters parallel and vertical to the electrodes are considered. It
is found that the relative orientation affects the transport
properties of these two devices completely differently. In the
Al(100)-Au$_7$-Al(100) device, the conductance and the current of
the parallel configuration are much larger than those in the
vertical configuration, while in the Al(100)-Ag$_{3}$-Al(100)
device, an opposite conclusion is obtained.

\end{abstract}
\pacs{85.65.+h, 73.63. -b, 36.40.-c \\[-5mm]}
\maketitle

The electron transport properties of molecules and clusters have
attracted tremendous interest in recent years since molecular
devices made from special kinds of molecules or clusters may find
important applications in future electronic circuits. \cite{1,2,3}
With the advantages of experimental techniques, for example,
scanning tunneling microscope (STM) and mechanically controllable
break junction (MCBJ), the measurement of current through nanoscale
systems is now allowed. An amount of interesting behaviors, such as
highly nonlinear $I$-$V$ characteristics, negative differential
resistance (NDR) and electric switching behavior, are found in
various systems such as organics, \cite{4} carbon nanotubes,
\cite{5} DNA, \cite{6,7} etc. A distinct difference between the
molecular devices and traditional conductors is that the transport
properties of the molecular devices depend not only on the
characteristics of the functional molecules or clusters themselves,
but also on many other factors, such as contact geometry and gate
voltage.\cite{8} In particular, in experiment, it is very difficult
to determine how the molecule or the cluster is put between the
electrodes.\cite{9} In other words, it is hard to know the details
about the contact geometry. In fact, in experiment, generally,
measurements are taken for many times and the average is performed
to obtain the conductance of a molecular conductor so that the
details of the contact geometry can be taken into consideration as
fully as possible. However, in theoretical simulation, it is easy to
design all kinds of possible contact geometries, for example, the
distance between the central cluster or molecule and the electrodes,
the orientation of the molecule relative to the electrodes, etc.

\begin{figure}[t]
\epsfig{file=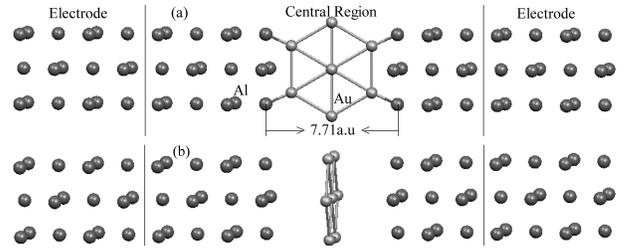, width=8cm} \caption{Model structure of a
two-probe system with Au$_7$ cluster coupled to two Al(100)
electrodes: (a) parallel configuration and (b) vertical
configuration.}
\end{figure}

\begin{figure}[t]
\epsfig{file=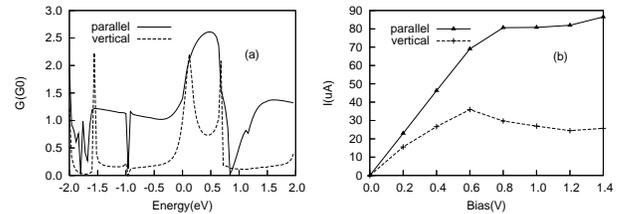, width=8cm} \caption{The conductance curves
(a) and $I$-$V$ curves (b) of the Au$_7$ system: the solid line
for the parallel configuration, and the dotted line for the
vertical configuration. The Fermi energy is set to zero (the same
for all the following figures)}
\end{figure}

 In this Letter, we study the effects of the contact geometry on the
transport properties by two model systems: Al(100)-Au$_7$-Al(100)
and Al(100)-Ag$_3$-Al(100), where both Au$_7$ and Ag$_3$ are stable
clusters with  planer structures. In a general measurement, the
distance between the electrodes is fixed while the orientation of
central molecule relative to the electrodes is undetermined.
Therefore, in our simulation, we will only investigate the effects
of the orientation of the cluster relative to the electrodes, with
the distance between the electrodes fixed. Particularly, in each
model system, two typical configurations, with the cluster plane
parallel and vertical to the electrodes, are studied.

Our calculations are performed using a recently developed
first-principles package TranSIESTA-C, which is based on the NEGF
technique. The TranSIESTA-C, as is implemented in the well tested
SIESTA method, is capable of fully self-consistently modeling the
electrical properties of nanoscale devices that consist of an atomic
scale system coupling with two semi-infinite electrodes as shown in
Fig.1. Details of the method and relevant references can be obtained
elsewhere\cite {10,13}.

\begin{figure}[t]
\epsfig{file=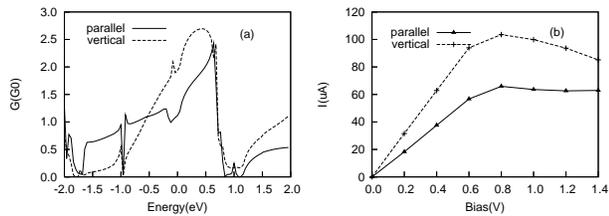, width=8cm} \caption{The conductance curve
(a) and $I$-$V$ curve (b) of the Ag$_3$ system: the solid line for
the parallel configuration, and the dotted line for the vertical
configuration.}
\end{figure}

\begin{figure}[t]
\epsfig{file=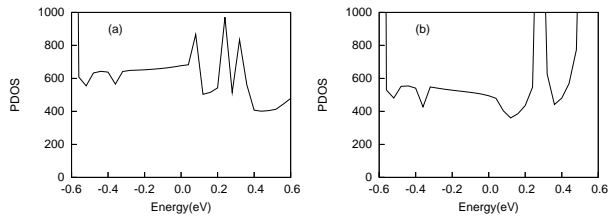, width=8cm} \caption{The PDOS of the two
configurations of the Au$_7$ systems at bias 1.2$V$: (a) for the
parallel configuration; (b) for vertical configuration.}
\end{figure}

 The model structure for our theoretical
analysis is illustrated in Fig.1. The Au$_7$ or Ag$_3$ cluster is
sandwiched between the two Al (100) electrodes with finite cross
section. The electrode is chosen from the perfect Al crystal along
the (100) direction, and the number of atoms in each atomic layer is
arranged as 5,4,5,4¡­ Four Al atomic layers (5, 4, 5, 4) are
selected for the electrode cell.  The noble metal cluster Au$_7$ or
Ag$_3$ together with four surface atomic layers in the left
electrode and three surface atomic layers in the right electrode is
selected as the central scattering region, as indicated by the two
vertical lines. These atomic layers in the central scattering region
are thick enough so that the perturbation effect from the cluster to
the electrodes is limited to these layers, and therefore the
remaining part of the electrodes can be treated as periodic systems.
The distance between the electrodes is fixed at $d=7.71 a.u.$ for
the Au$_7$ case, and 5.85 $a.u.$ for Ag$_3$ case. The average
nearest-neighbor distance of Au-Au bond is 2.72 $a.u.$; it's $2.67
a.u.$ for Ag-Ag bond \cite{14}. In our calculation, the convergence
criterion for the Hamiltonian, charge density, and band structure
energy is 10$^4$ and the atomic cores are described by
normconserving persudopotentials.

\begin{figure}[t]
\epsfig{file=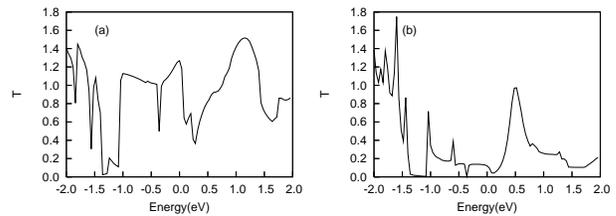, width=8cm} \caption{(The transmission
coefficient of the two configurations of Au$_7$ as a function of
energy at the applied bias voltage of 1.2$V$: (a) for the parallel
configuration; (b) for vertical configuration. }
\end{figure}

\begin{figure}[t]
\epsfig{file=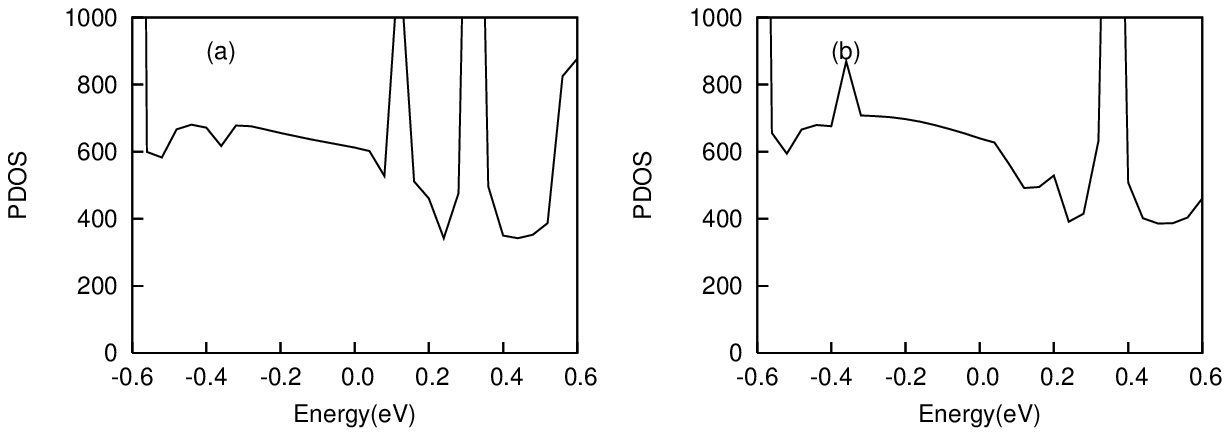, width=8cm} \caption{The PDOS of the two
configurations of the Ag$_3$ systems at bias 1.2$V$: (a) for the
parallel configuration; (b) for vertical configuration.}
\end{figure}

\begin{figure}[t]
\epsfig{file=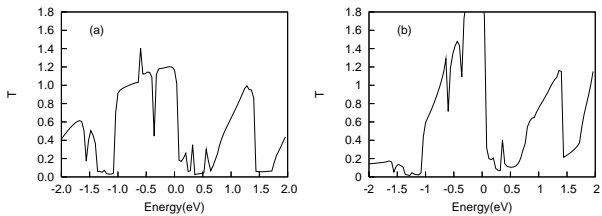, width=8cm} \caption{The transmission
coefficient of the two configurations of Ag$_3$ as a function of
energy at the applied bias voltage of 1.2$V$: (a) for the parallel
configuration; (b) for vertical configuration.}
\end{figure}

The conductance curves of the Au$_7$ cluster in two different
configurations are presented in Fig.2(a). From the figure, it is
found that the equilibrium conductance in the parallel case is much
bigger than that in the vertical case. To investigate its
non-equilibrium transport properties, the $I$-$V$ curves of the two
configurations of Au$_7$ cluster are also shown in Fig.2(b).
Obviously, the current of the parallel configuration is larger than
that of the vertical configuration over the given bias range. With
the increase of the bias, the difference in the current of the two
configurations is getting bigger and bigger, and the current of the
parallel configuration is as nearly 4 times as that of the vertical
one at 1.4$V$. While for Ag$_3$ cluster, the equilibrium conductance
(shown in Fig.3(a) and the current (shown in Fig.3(b) of the
parallel configuration are smaller than those of the vertical one.

The different effects of the relative orientation of the clusters on
the transport properties in these two devices can be understood by
studying the coupling between the electrode and the molecule via the
projected density of states (PDOS) because it can give us
information on how strongly the molecule couples with the
electrodes\cite{15}. Without loss of generality, in both devices,
the PDOS of the two configurations under bias 1.2$V$ as an example
will be considered.

The PDOS and the transmission coefficient as a function of energy of
the two configurations of Au$_7$ cluster are shown in Figs.4 and 5.
It can be seen in Fig.4 that near the Fermi energy, the PDOS of the
parallel configuration is much larger than that of the vertical
configuration. Corresponding to Fig.4, the transmission coefficient
of the parallel configuration also takes larger values near the
Fermi energy than that of the vertical configuration. Then we
conclude that in the parallel configuration the Au$_7$ cluster has
strong coupling with electrodes via Au-Al bond, which makes the
electrons with an energy near the Fermi level transmit through the
cluster easily, thus give rise to a very strong transmission and
large conductance. But no Au-Al bond is formed and two vacuum
potential barriers exist between the electrodes and Au$_7$ in the
vertical configuration. Therefore the coupling between the cluster
and the electrodes is rather weak, which results in much smaller
conductance for the vertical configuration. The current $I$ is
calculated by the equation
$I=\frac{2e}{h}\int_{-\frac{V}{2}}^{\frac{V}{2}}dET(E,V_{b})$, where
the transmission $T(E,V)$ is a function of the bias $V$ and the
incident electron energy E. At the applied bias voltage of 1.2$V$,
the transmission coefficient of the vertical configuration takes
much smaller values than that of the parallel configuration at the
energy range [-0.6,0.6] $eV$. Therefore, the current of the parallel
configuration has much larger value than that of the vertical
configuration.

The corresponding PDOS and transmission coefficient for the two
configurations of the Ag$_3$ case at bias 1.2$V$ are shown in Fig.
6. In contrast, the PDOS of the parallel configuration is smaller
than that of the vertical configuration. The transmission
coefficient of the parallel configuration also takes smaller values
near the Fermi energy than that of the vertical configuration (see
Fig. 7). Thus the current and the conductance of the parallel
configuration are smaller than those of the vertical configuration
in the Ag$_3$ cluster. The reason is that, when rotating the Ag$_3$
cluster from a parallel configuration to a vertical one, no vacuum
barrier is formed between the electrodes and the cluster.
Furthermore, the coupling for the vertical configuration gets
stronger than that for the parallel configuration, since electrons
can transmit through the three Ag atoms in a parallel way, not a
series way now.

The change of the coupling strength when the cluster is rotated can
also be displayed by the charge transfer between the electrodes and
the cluster.\cite{16,17} For the Au$_7$ case, in the parallel
configuration, due to the stronger coupling, there is a charge
transfer of 0.92$e$ from the Au$_7$ cluster to the electrodes, while
in the vertical configuration, it is just 0.61$e$.  In contrast, for
the Ag$_3$ case, 0.87$e$ charge is transferred from the cluster to
the electrodes due to weaker coupling for the parallel configuration
while 0.92$e$ due to stronger coupling for the vertical
configuration.

In summary, a first-principles method is used to study the effects
of the relative orientation of the molecules on the electron
transport in molecular devices. Two molecular devices, with the
planer Au$_7$ and Ag$_3$ cluster sandwiched between the Al(100)
electrodes are investigated. We find that the relative orientation
affects the transport properties of these two devices completely
differently. In the Al(100)-Au$_7$-Al(100) device, the conductance
and the current of the parallel configuration are much bigger than
those in the vertical configuration, while in the
Al(100)-Ag$_3$-Al(100) device, it is just the opposite case. The
different effects of the relative orientation of the clusters on the
transport properties in the two molecular devices are attributed to
the different coupling in the molecule/electrode contacts. In the
vertical configuration of Au$_7$, two vacuum barriers are formed
between the electrodes and the cluster, thus the coupling becomes
much weaker when the Au$_7$ cluster is rotated from the parallel
configuration to the vertical one. However, for the Ag$_3$ case, on
one hand, no barrier is formed when the cluster is rotated from the
parallel configuration to the vertical one; on the other hand, all
the three atoms of the cluster play the role of a bridge in a
parallel style and strengthen the coupling between the electrodes
and the cluster. Consequently, the conductance and the current are
also strengthened in the vertical configuration of the Ag$_3$ case.

This work was supported by the NSFC under Grant No. 10404010, and
Postgraduate Innovation Foundation of Jiangxi Normal University.

\end{document}